\documentclass[journal,comsoc]{IEEEtran}
\usepackage[T1]{fontenc}

\ifCLASSINFOpdf
   \usepackage[pdftex]{graphicx}
   \graphicspath{{../pdf/}{../jpeg/}}
\else
   \usepackage[dvips]{graphicx}
   \graphicspath{{../eps/}}
   \DeclareGraphicsExtensions{.eps}
\fi
\usepackage{amsmath}
\usepackage[cmintegrals]{newtxmath}
\usepackage[tight,footnotesize]{subfigure}

\begin{document}
%
\title{Finding the Sweet Spot for Frame Aggregation in 802.11 WLANs}
%
%
%

\author{Jose~Saldana,~\IEEEmembership{Senior Member,~IEEE,}
		Omer Topal,        
        Jos\'e~Ruiz-Mas,
        and~Juli\'an~Fern\'andez-Navajas
\thanks{J. Saldana, J. Ruiz-Mas and J. Fern\'andez-Navajas are with CeNIT Group - Aragon Institute of Engineering Research (I3A), EINA, University of Zaragoza, Spain. Email: \{jsaldana, jruiz, navajas\}@unizar.es. O. Topal is with Research and Development Department, AirTies, Turkey. Email:
omer.topal@airties.com}
\thanks{This work has been partially financed by the European Social Fund and Government of Aragon CeNIT Research Group T31 20R, and by UZ2019-TEC-03, Univ. Zaragoza}
\thanks{Manuscript received Jun 26, 2020; revised Nov 27, 2020.}}

%
%

\markboth{Journal of \LaTeX\ Class Files,~Vol.~14, No.~8, August~2015}%
{Shell \MakeLowercase{\textit{et al.}}: Bare Demo of IEEEtran.cls for IEEE Communications Society Journals}
%



\maketitle

\begin{abstract}
This letter proposes an algorithm for the dynamic tuning of the maximum size of aggregated frames in 802.11 WLANs. Traffic flows with opposed requirements may coexist in these networks: traditional services as web browsing or file download that need high throughput, and services with real-time requirements that need low latency. The proposed algorithm allows the network manager to find an optimal balance (\textit{i.e.} the ``sweet spot'' between throughput and latency: a ``delay budget'' can be assigned to real-time flows, with the objective of keeping the latency as close as possible to that budget, while penalizing the throughput of traditional services as little as possible.
\end{abstract}


%
\IEEEpeerreviewmaketitle

\section{Introduction}
%
%
%
%
\IEEEPARstart{F}{rame} 
rame aggregation can provide a significant efficiency improvement in 802.11 networks \cite{Ginzburg}. It was included for the first time in 802.11n, and it has been present in all the subsequent versions of the standard. There are two different aggregation schemes: A-MSDU (Aggregated MAC Service Data Unit) sends a number of MSDUs sharing a PHY header; in A-MPDU (Aggregation MAC Protocol Data Unit) scheme, a number of MPDUs are sent back-to-back, once access to the channel has been obtained. The latter scheme is widely used, and 802.11ac made it compulsory for all the frames.

The traffic mix that can be found nowadays in these WLANs combines flows of different nature, with diverse requirements. Traditional services (web browsing, file download) that just need throughput, share the network with other flows for which latency is critical, and typically consist of small packets \cite{Xie} with different delay constraints \cite{Suznjevic}.

As an example, in \cite{Lv}, a novel medium access control (MAC) scheme for hospital e-health systems was proposed, to reduce the latency of delay-sensitive healthcare traffic. The delay constraint was extremely low: 200 $\mu$s. Other studies consider that ultra-low latency means that the delay is looser, in the order of some milliseconds (\textit{i.e.} $\leq$ 5 ms) \cite{Alameddine}. Another study looking for ultra-low latency in WLANs \cite{Grazia} puts the target in 1 ms, considering emerging applications as vehicular or ultra-reliable communications. Finally, in \cite{Seytnazarov}, a QoS-aware adaptive scheduler was proposed, which was able to keep VoIP delay under a limit, based on cross-layer optimization, using certain RTP (Real-Time Protocol) features.

When a STA sends or receives a big aggregate frame, the rest of the STAs sharing the same AP have to wait, and this results in an increased delay and jitter for them. Therefore, frame aggregation has a double effect: its efficiency increase is highly beneficial for bandwidth-hungry flows, but it comes at the cost of undesired delays for real-time ones.

In order to mitigate these bad effects for real-time flows, the authors of \cite{Saldana} studied a scenario in which a central controller could enable or disable the aggregation features of a set of Wi-Fi APs belonging to the same local deployment. The mechanism gathered information from the network, and disabled aggregation in an AP whenever a flow with real-time requirements became associated to it. As soon as it left the AP (due to user movement), aggregation could be enabled again, as long as there were no more STAs generating that kind of traffic. As an alternative, the study also considered the possibility of permanently configuring a reduced value of the maximum size of the A-MPDU, which also provided some advantages. In \cite{Coronado} the use of Machine Learning (ML) techniques was proposed for adaptively adjusting the maximum A-MSDU length on a per-user basis. The paper was mainly focused on reducing packet loss rate and increasing channel occupation.

In this research context, the main contribution of the present paper is the proposal of an algorithm that dynamically tunes the value of the maximum A-MPDU in order to meet a “delay threshold,” defined as the maximum latency tolerated by the real-time flows present in the WLAN. It can be considered as the take of the total “delay budget” that can be “spent” in the Wi-Fi network. As we will next explain, different methods can be employed by the algorithm.

The difference with respect to \cite{Saldana} is that we do not follow an on/off approach for the aggregation, but a real-time adaptation that will allow us to find the “sweet spot” of the trade-off between throughput and latency explored in that paper.

There are two differences with respect to \cite{Coronado} first, we are not considering A-MSDU (length up to 3,839 bytes), but A-MPDU (length up to 65,535 bytes), which is more widely used and may result in a much higher delay for real-time flows; second, our objective is not to achieve the maximum throughput, but to achieve the optimal balance between the users of real-time applications and ``traditional'' ones that share the same network.

The objective of the proposed algorithm (we will refer to it as the \textit{tuning} algorithm from now on) is to maximize the throughput of the TCP flows, while granting that the latency of real-time STAs is below the desired threshold.

It behaves as follows: at the beginning, the APs start with the maximum A-MPDU value, \textit{i.e.}, they aggregate as many frames as possible. The flows with real-time requirements are periodically monitored, and their maximum delay is obtained at each AP. If this delay is above a threshold, the A-MPDU size is reduced. If the delay is below the threshold, the A-MPDU limit is increased until it reaches the maximum.

Different methods for increasing and decreasing the limit can be employed:

\begin{itemize}
  \item a linear increase / decrease with a fixed-size step;
  \item a geometrical increase / decrease, using a coefficient for obtaining the new value;
  \item a \textit{drastic} increase / decrease, \textit{i.e.} as soon as the value exceeds the threshold, the aggregation is enabled / disabled.
\end{itemize}

Obviously, a different method can be employed for increasing and decreasing the value. This results into a total of 9 possible combinations, among which the following ones have been chosen as significant (the concrete values of the increasing / decreasing steps have been found empirically as the ones that provide the best results, as we will see in Section ~\ref{sec:tests}):
 
\begin{itemize}
  \item Method \#1. Linear increase and linear decrease: if the latency is above the objective, the maximum A-MPDU value is decreased by a fixed value (3,000 bytes). If it is below, the value is increased by the same value.

  \item Method \#2. Geometric decrease and increase: if the latency is above the limit, the A-MPDU is reduced by a factor of 0.618; if it is below, it is multiplied by its inverse (1.618).
  \item Method \#3. Instantaneous drastic reduction to the minimum, linear increase: if the latency is above the latency budget, the A-MPDU is reduced to the minimum. If it is below, the A-MPDU is increased by 6,000 bytes.
  
  \item Method \#4. Linear decrease, and instantaneous \textit{drastic} rise to the maximum: if the latency is below the latency budget, the A-MPDU is increased to the maximum. If it is above, it is decreased by 6,000 bytes.
\end{itemize}
 
As a baseline, we will use the method in which aggregation is just activated or de-activated \cite{Saldana} (we will call it the \textit{disable} algorithm, and it is in fact a particular case including a \textit{drastic} increase and a \textit{drastic} decrease).

In the next section we present the test setup, and in Section~\ref{sec:tests} the results are presented and compared. The paper ends with the Conclusions.

\section{Test Setup}
\label{sec:setup}

\begin{table}[!t]
\renewcommand{\arraystretch}{1.3}
\caption{Simulation Parameters}
\label{table_1}
\centering
\begin{tabular}{c|c||c|c}
\hline
\textbf{ns3 version} & ns-3.30.1 & \textbf{UDP pkt size} & 60 bytes\\
\hline

\textbf{Mobility model} & Random Waypoint & \textbf{UDP rate} & 50 pps\\
\hline

\textbf{Walking speed} & 1.5 m/s with & \textbf{TCP pkt size} & 1,500 bytes\\
& pause time 2 s & & \\
\hline

\textbf{WiFi model}& \textit{SpectrumWifiPhy} & \textbf{TCP variant} & New Reno\\
& with \textit{MultiModel} & & \\
& \textit{SpectrumChannel} & & \\
\hline

\textbf{Channels} & 36 to 128 (20 & \textbf{Simulation} & 60 s\\
& MHz channels) &\textbf{time}& \\
\hline

\textbf{WiFi phy std}&802.11ac &\textbf{RTS/CTS}&Enabled\\
\hline

\textbf{WiFi rate}& \textit{Idealwifi} & \textbf{Inter-AP} & 50 m (25 m\\
\textbf{control model}& Manager & \textbf{distance} &to the border)\\
\hline

\textbf{Propagation} & \textit{FriisSpectrum} & \textbf{Short guard} & Not enabled\\
\textbf{loss model} & & & \\
\hline

\textbf{Error rate} & \textit{NistErrorRateModel} & \textbf{EDCA} & Enabled\\
\textbf{model} & &\textbf{priorities} & \\
\hline

\textbf{Delay moni-} & \textit{250 ms} & \textbf{A-MPDU} & 1,600 to\\
\textbf{toring period} & &\textbf{range} & 65,535 bytes\\
\hline
\end{tabular}
\end{table}

The tests have been performed using ns3 simulations of an 802.11 scenario with a number of APs and a central controller. As a starting point we used the code released by the authors of \cite{Saldana}, in which we have included new features as \textit{e.g.} 802.11ac. On behalf of accuracy, we have also used the next models instead of the original ones:

\begin{itemize}
  \item \textit{Idealwifi}, an SNR-based rate manager, in which a transmitter learns the SNR information at the receiver by an out-band channel. It directly manages the rates and also considers the bit error rates (BER) during transmissions.

  \item \textit{FriisSpectrum} propagation loss model is used to take the frequency into account in the tests
  \footnote{See \mbox{https://www.nsnam.org/doxygen/classns3\_1\_1\_friis\_spectrum} \mbox{\_propagation\_loss\_model.html}}. It is important, since it directly impacts the RSSI and hence system performance.

  \item \textit{Nist} error rate model, which has a performance closer to real behavior than other ones (\textit{e.g. YansWifi}) \cite{Pei}.
\end{itemize}

Table \ref{table_1} summarizes the parameters employed. The code used in the presented tests has been released (see https://github.com/wifi-sweet-spot/ns3).

In all the simulations, two kinds of STAs share the scenario: \textit{a)} ``TCP STAs'' are performing a file download, and \textit{b)} ``UDP STAs'' are running an application that sends small packets with real-time requirements (see Table \ref{table_1}). In all cases, 2\textit{N} STAs will be present: \textit{N} TCP STAs and \textit{N} UDP STAs. These applications have opposite target Key Performance Indicators (KPIs): TCP STAs want high throughput whereas UDP STAs’ main objective is to keep the latency low.

An interval of 250 ms has been used as the time between adjustments of the maximum A-MPDU size. This value was found empirically, as it presented a good trade-off between a fast adjustment (it correctly keeps track of the evolution of the latency) and a low computation overhead.

\section{Tests and Results}
\label{sec:tests}

We will first present a scenario with a single AP and two STAs (one TCP and one UDP), which will allow us to qualitatively examine the behavior of the proposed algorithm, and the ones used for comparison purposes. Then, we will use a bigger scenario with 4x4 APs and a variable number of STAs, in order to quantitatively measure the overall parameters (latency and throughput) with more accuracy.

\subsection{Scenario with a single AP and two STAs}

In this subsection we use a 50x50 meter scenario that includes a single AP shared by two Wi-Fi STAs (\textit{N}=1). We present three experiments: \textit{a) no aggregation; b) aggregation always active; c) tuning algorithm, method} \# 1. They will let us observe the behavior of the A-MPDU adaptation in detail.

Note that the \textit{disable} algorithm \cite{Saldana} would produce the same results as the \textit{no aggregation} case: since a TCP STA is always connected to the AP, aggregation in the AP would in fact be always disabled.

\begin{figure}
\centering
\includegraphics[width=2.7in]{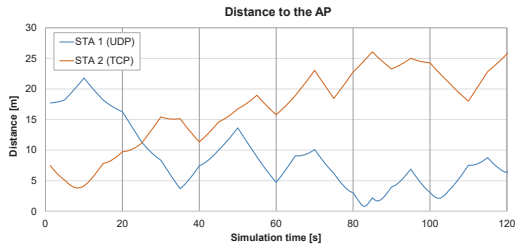}
\caption{Distance of the STAs to the AP.}
\label{figure1}
\vspace*{-0.2in}
\end{figure}

The same random seed has been used in the three experiments, in order to have the same mobility patterns. In Fig. \ref{figure1} (valid for \textit{a}, \textit{b} and \textit{c}), the distance from the STAs to the AP is shown. It can be seen that the TCP STA gets very close to the AP in \textit{t} = 8 s. From that moment, it moves away from it, while the UDP STA gets closer.

\begin{figure}[!t]
\centering
{
\subfigure[]{\includegraphics[width=3.0in]{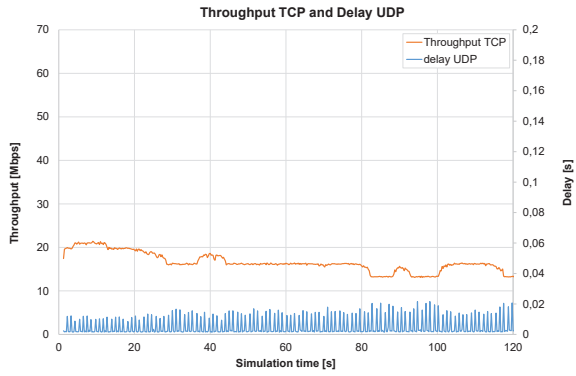}
\label{figure2a}}
\subfigure[]{\includegraphics[width=3.0in]{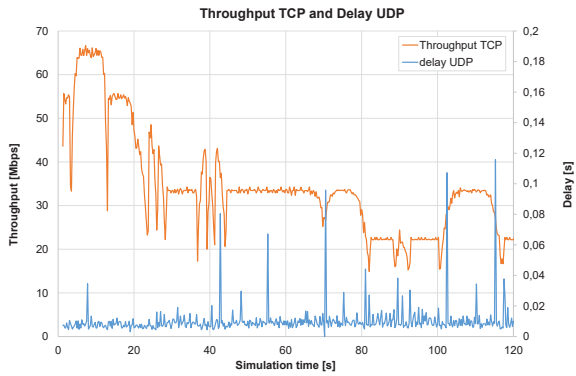}
\label{figure2b}}
\subfigure[]{\includegraphics[width=3.0in]{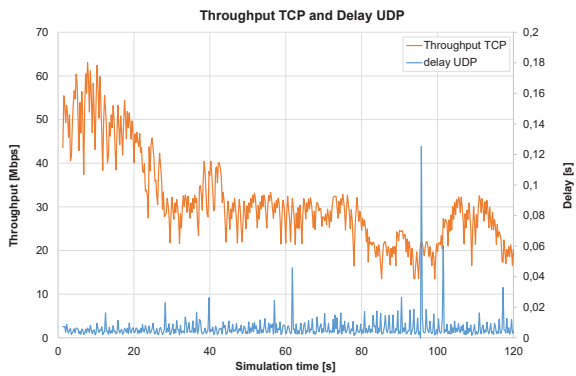}
\label{figure2c}}
\caption{Single AP scenario: TCP throughput and UDP real-time delay: \textit{a)} no aggregation; \textit{b)} aggregation enabled with a constant value of A-MPDU size of 65 KB; \textit{c}) the \textit{tuning} algorithm (\textit{method} \#1) limits the maximum A-MPDU value.}
\label{figure2}
}
\vspace*{-0.2in}
\end{figure}

Fig. \ref{figure2} shows the results of the \textit{no aggregation} experiment. The highest TCP throughput is obtained around \textit{t} = 8 s, and the delay is always below 20 ms (the average delay is 4.7 ms, and the average throughput is 16.6 Mbps).

In the \textit{aggregation always active} case, (Fig. \ref{figure2b}), the throughput obtained by the TCP STA is much higher (34.5 Mbps  average), but this comes at the cost of a higher latency for the UDP STA (10 ms average) . Some latency spikes of the UDP flow appear when the distance of the TCP STA grows, since its rate gets reduced and A-MPDUs can be up to 65 KB.

\begin{figure}
\centering
\includegraphics[width=2.7in]{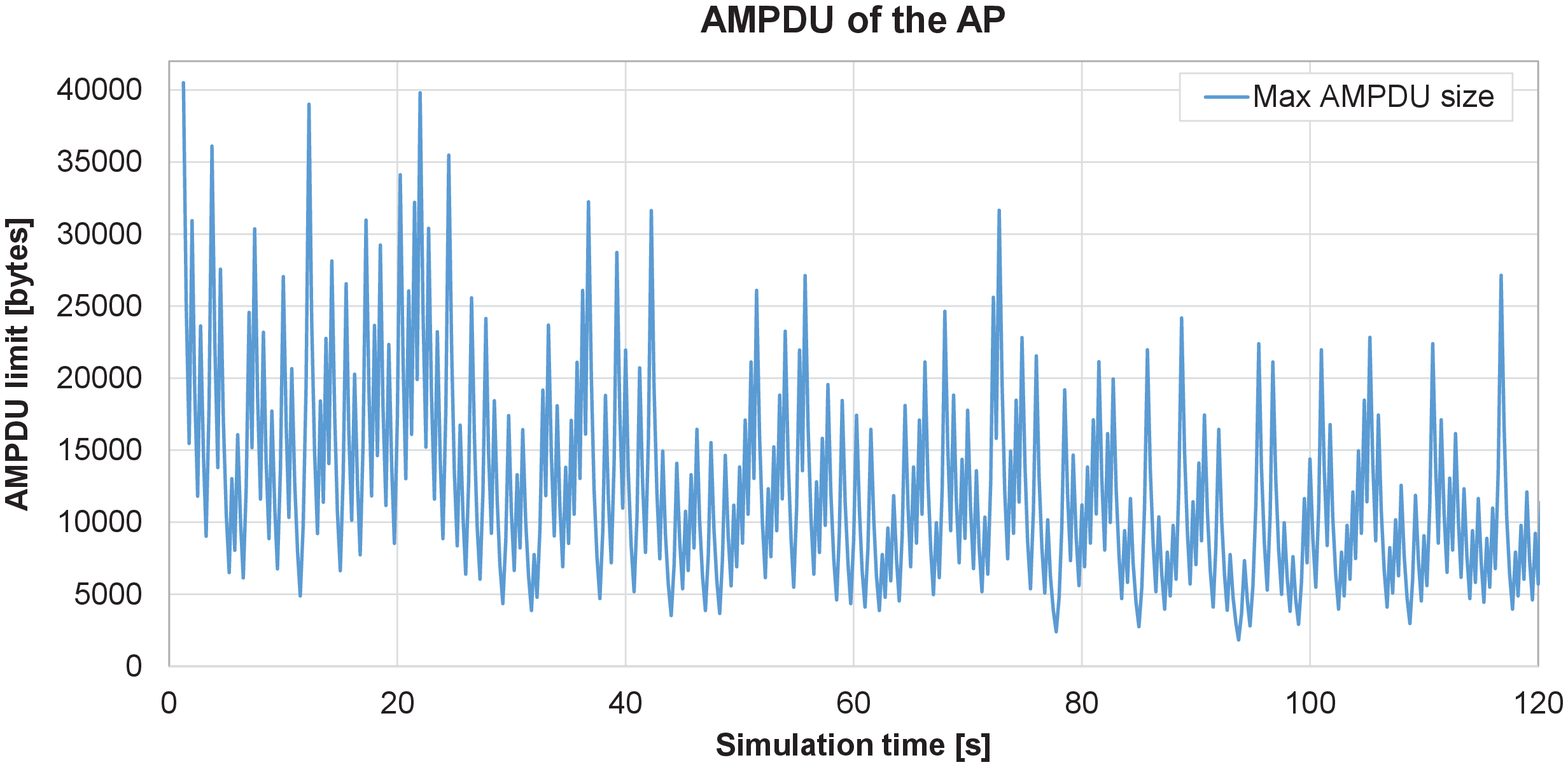}
\caption{Single AP scenario with \textit{method} \#1: evolution of the maximum A-MPDU value.}
\label{figure3}
\vspace*{-0.4in}
\end{figure}

Finally, if the \textit{tuning} algorithm (\textit{method} \#1) is employed (Fig. \ref{figure2c}), we can still maintain a high throughput for the TCP STA (31.3 Mbps , \textit{i.e.} only 10 percent reduction with respect to the \textit{aggregation always active}), whereas the UDP latency presents a behavior somewhat similar to the one obtained without aggregation. Its average delay is now 6.2 ms (25 percent increase with respect to \textit{no aggregation}). Fig. \ref{figure3} shows the evolution of the A-MPDU maximum size as the algorithm adapts it as a function of the UDP latency.

The trade-off appears clearly: the system sacrifices some throughput of the TCP flow in order to grant a low delay for the real-time one. The \textit{tuning} algorithm allows us to find and set this ``sweet spot,'' giving us the possibility to tune the trade-off between maximum throughput for TCP and minimum latency for UDP.

In the next subsection we will measure this in detail, with a higher number of realizations, more APs and different numbers of STAs sharing the scenario.

\subsection{Scenario with a number of APs}

In order to capture the effect of the proposed \textit{tuning} algorithm, this section presents averaged results obtained in a scenario with a higher number of APs and STAs: it consists of a set of 16 APs, deployed in a 4x4 array with 50 meters between them, and 25 m to the border (total 200 x 200 m). A different number of UDP and TCP STAs (from 4 to 100, \textit{i.e. N} = 2 to 50) move through the scenario while running a TCP or a UDP application.

The four methods are compared between them, and also versus \textit{no aggregation}, \textit{aggregation always active}, and the \textit{disable} algorithm. Fig. 4 shows the total throughput obtained by the TCP STAs, and Fig. \ref{figure5} presents the average latency of the UDP STAs. The results are the average of 15 realizations of 60 seconds each (50 realizations were required for the \textit{aggregation always active} scenario). 95\% confidence intervals are provided.

It can be observed that the \textit{no aggregation} policy adds very low latency, but it is always the one with the lowest throughput: this could be expected, since the transmission of every single frame requires the use of the CSMA/CA mechanism.

In contrast, the \textit{aggregation always active} experiment is the one that achieves the highest throughput (which always doubles the one achieved by means of \textit{no aggregation}). However, this comes at the cost of a significantly higher latency (in some cases even doubled).

\begin{figure}
\centering
\includegraphics[width=\linewidth]{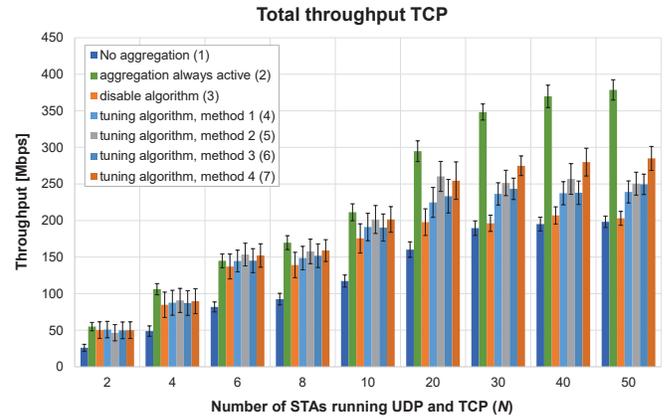}
\caption{Scenario with 16 APs: total throughput of the TCP STAs.}
\label{figure4}
\vspace*{-0.2in}
\end{figure}

\begin{figure}
\centering
\includegraphics[width=\linewidth]{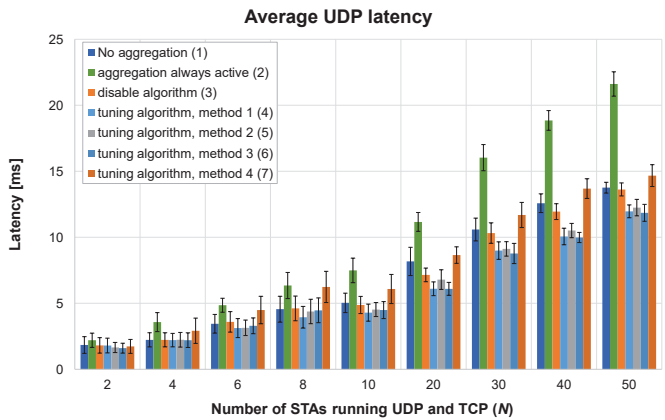}
\caption{Scenario with 16 APs: average latency of the UDP STAs.}
\label{figure5}
\vspace*{-0.2in}
\end{figure}

The test of the \textit{disable} algorithm reports the same latency as the one in the \textit{no aggregation} one: this could be expected: as soon as a UDP STA gets associated, the AP disables the aggregation, so this STA always \textit{sees} a non-aggregating AP. however, the throughput of the TCP STAs gets reduced significantly (between 52 and 81 percent ).

\begin{figure}
\centering
\includegraphics[width=\linewidth]{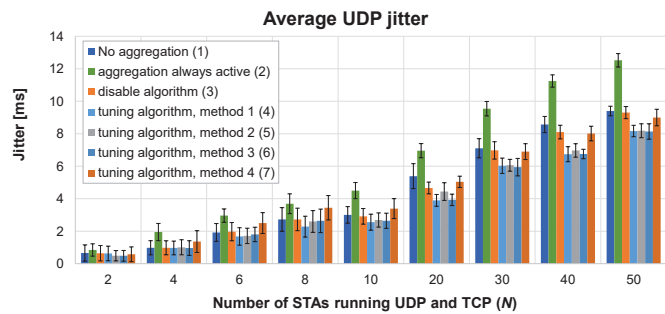}
\caption{Scenario with 16 APs: average jitter of the UDP STAs.}
\label{figure6}
\vspace*{-0.2in}
\end{figure}

Finally, if the proposed \textit{tuning} algorithm is used, it can be seen that the latency \textit{vs} throughput trade-off can be adjusted in order to find the most convenient point: the throughput (Fig. \ref{figure4}) is now much closer to the \textit{aggregation always active} case, while the latency (Fig. \ref{figure5}) is very similar to the one achieved by \textit{no aggregation} or the \textit{disable} algorithm. Fig. \ref{figure6} shows the jitter of the UDP traffic, in which the same tendency can be observed: it is very similar in all cases, except for the \textit{aggregation always active} one, in which the jitter is increased by roughly 70 \%.

\begin{figure}
\centering
\includegraphics[width=\linewidth]{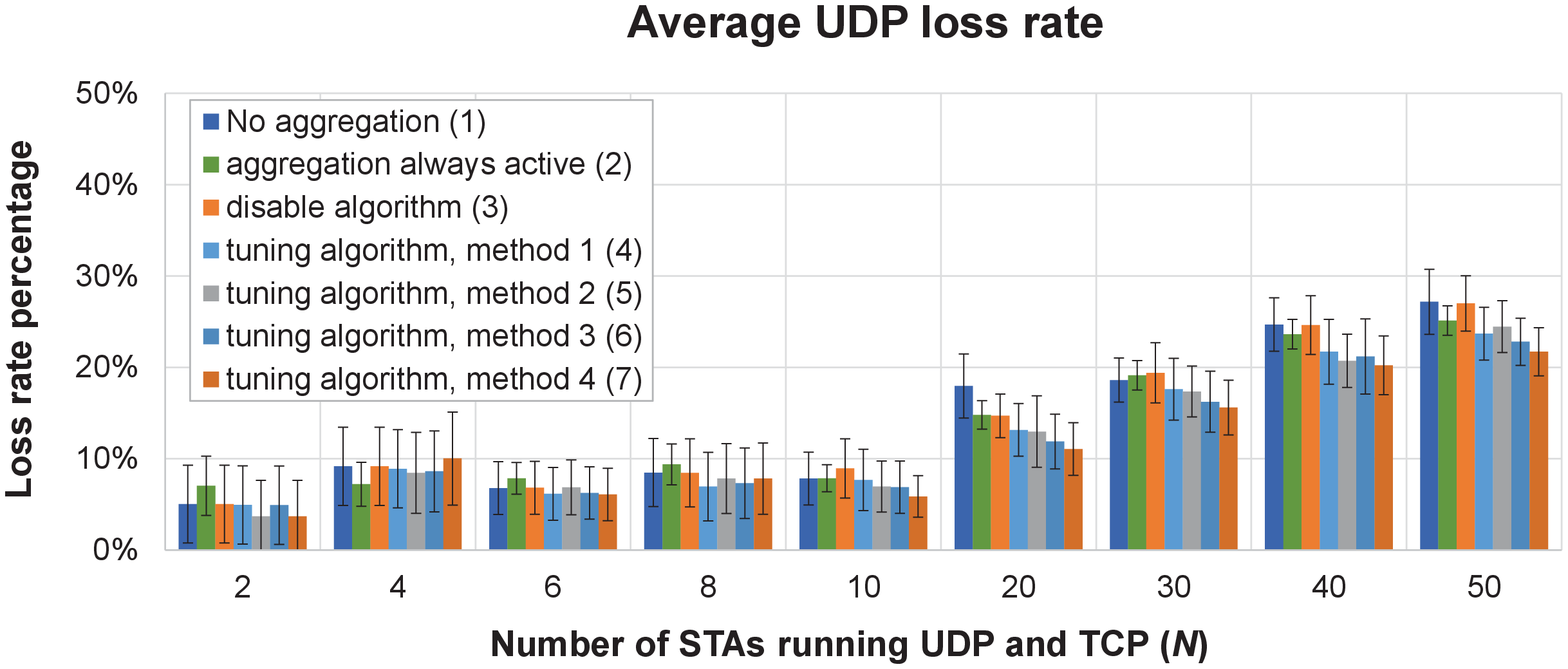}
\caption{Scenario with 16 APs: loss rate of the UDP STAs.}
\label{figure7}
\vspace*{-0.15in}
\end{figure}

As far as loss rate is concerned (Fig. \ref{figure7}), no special impact is observed if the different methods are employed: the maximum difference between them is 7\% . In addition, the loss rate is always below the values obtained by the \textit{no-aggregation} (14\% below on average) and \textit{aggregation always active} (18\% below on average) cases.

Regarding the differences between the four methods, \#4 is the one that grants a highest throughput for the TCP STAs. The cause is the fact of having a more aggressive rise when the latency is below the limit. However, this comes with a penalty in terms of latency for the UDP flows. To a lesser extent, something similar happens with \textit{method} \#2 (geometric increase and decrease), and \#3, which also has a fast rise in steps of 6,000 bytes and a drastic reduction.

At the same time, \textit{method} \#1 is the one with the best results in terms of latency for the UDP flows. The results obtained with \textit{method} \#3 are very similar to those obtained with \#1, with a slight throughput reduction caused by the aggressive policy that reduces the A-MPDU size to the minimum whenever the latency limit is reached.

All in all, it seems clear that the use of a soft increase and decrease draws better results in terms of latency, but it slightly reduces the throughput. This is quite noticeable for \textit{method} \#1 (linear) and also for \#3 (drastic reduction). In contrast, the aggressive rise of \textit{method} \#4 is better for bandwidth-hungry flows and worse for real-time ones.

\begin{figure}[!t]
\centering
{
\subfigure[]{\includegraphics[width=.47\linewidth]{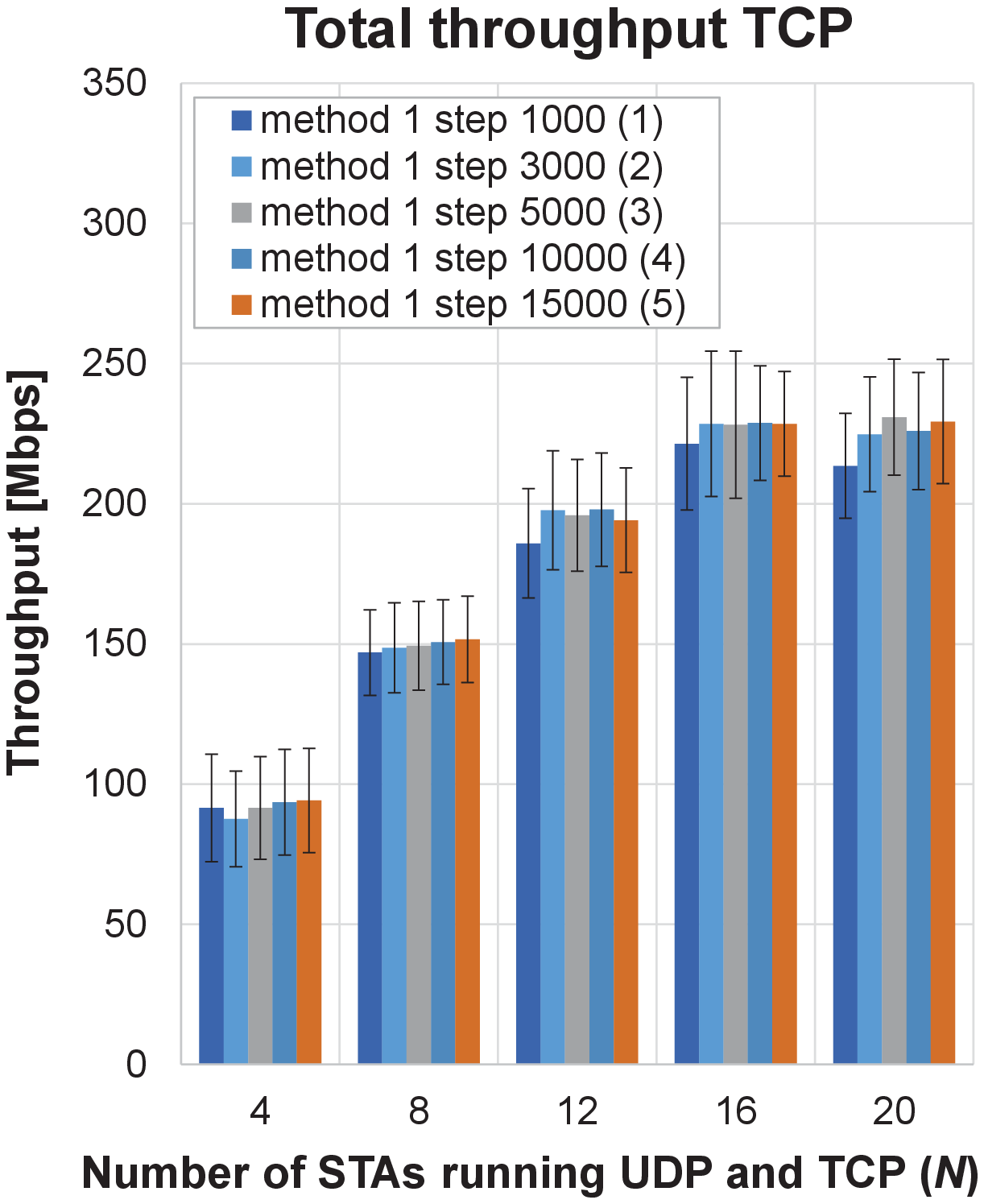}
\label{figure8a}}
\subfigure[]{\includegraphics[width=.47\linewidth]{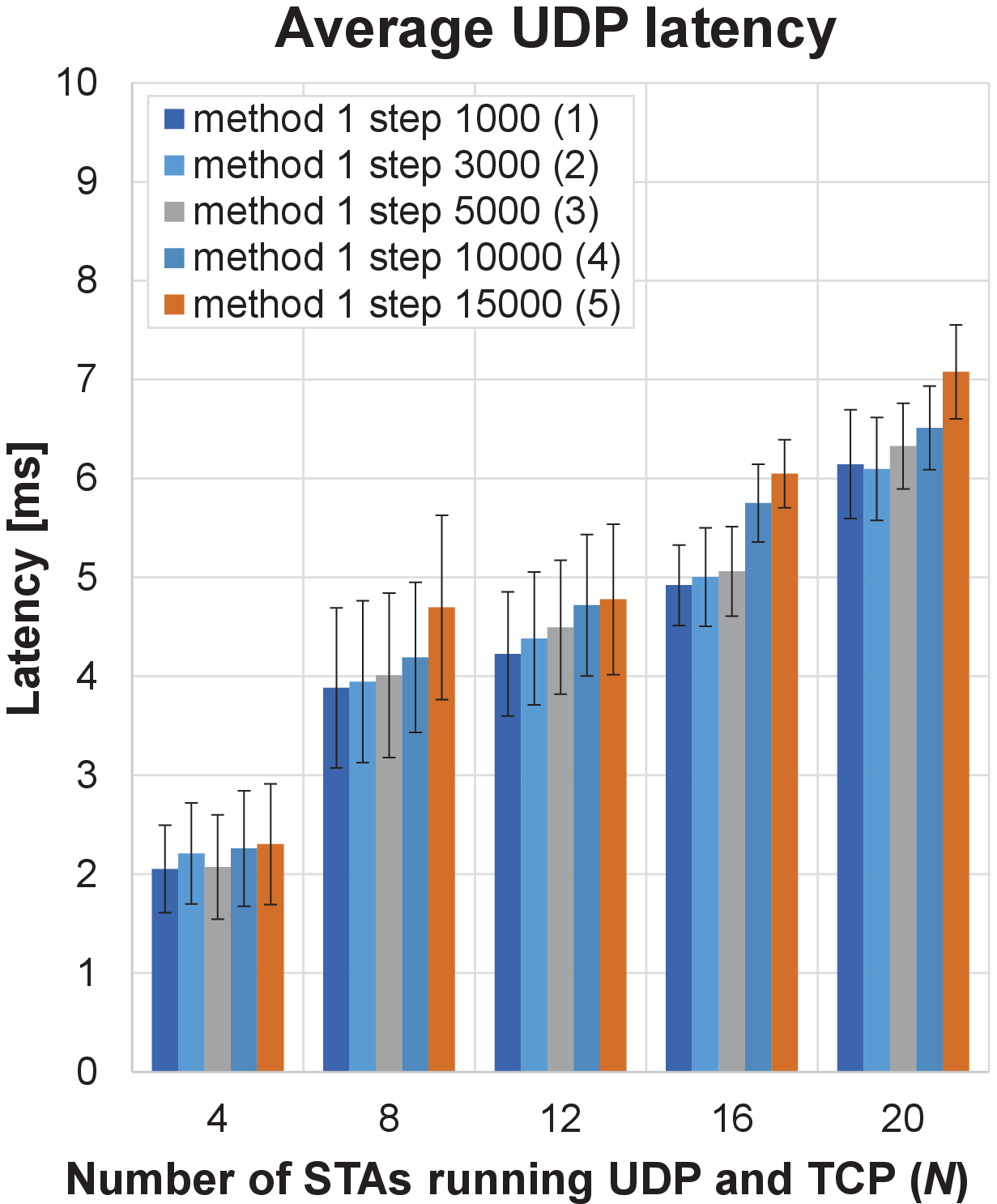}
\label{figure8b}}
\caption{Scenario with 16 APs: comparison for \textit{method} \#1 with different values of the step: \textit{a)} throughput; \textit{b)} latency.}
\label{figure8}
}
\vspace*{-0.15in}
\end{figure}

Fig. \ref{figure8} presents some of the tests that have been run in order to find the best value of the increasing / decreasing steps. In this case, \textit{method} \#1 with five different values for the step (ranging from 1,000 to 15,000) is compared. It can be seen that high values of the step are not convenient: with 10,000 and 15,000 we obtain a higher latency, but the throughput is not increased. In contrast, a value of 1,000 results in a lower throughput with the same latency than 3,000 or 5,000, which are the best values for the increasing / decreasing step.

\begin{figure}[!t]
\centering
{
\subfigure[]{\includegraphics[width=.47\linewidth]{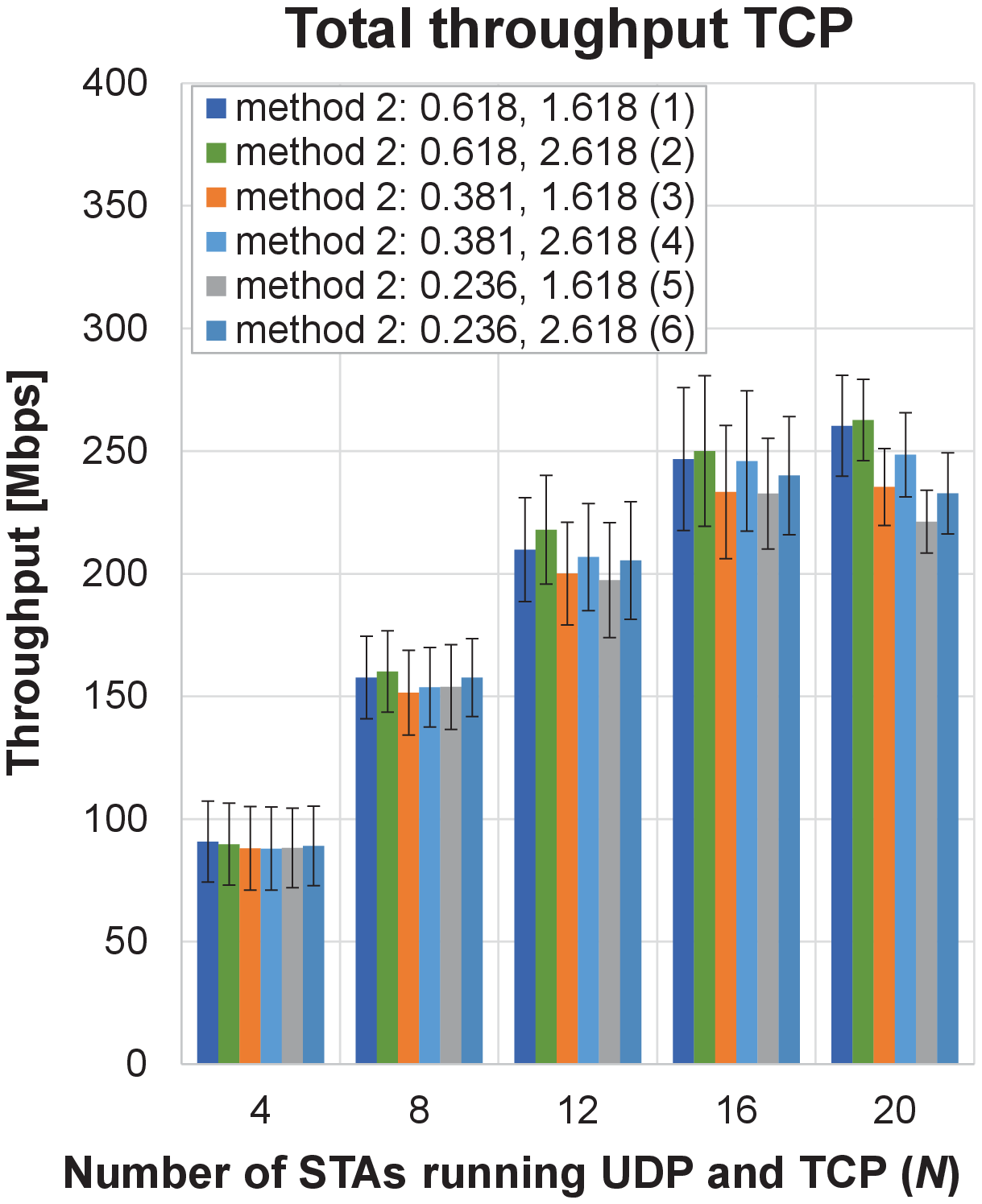}
\label{figure9a}}
\subfigure[]{\includegraphics[width=.47\linewidth]{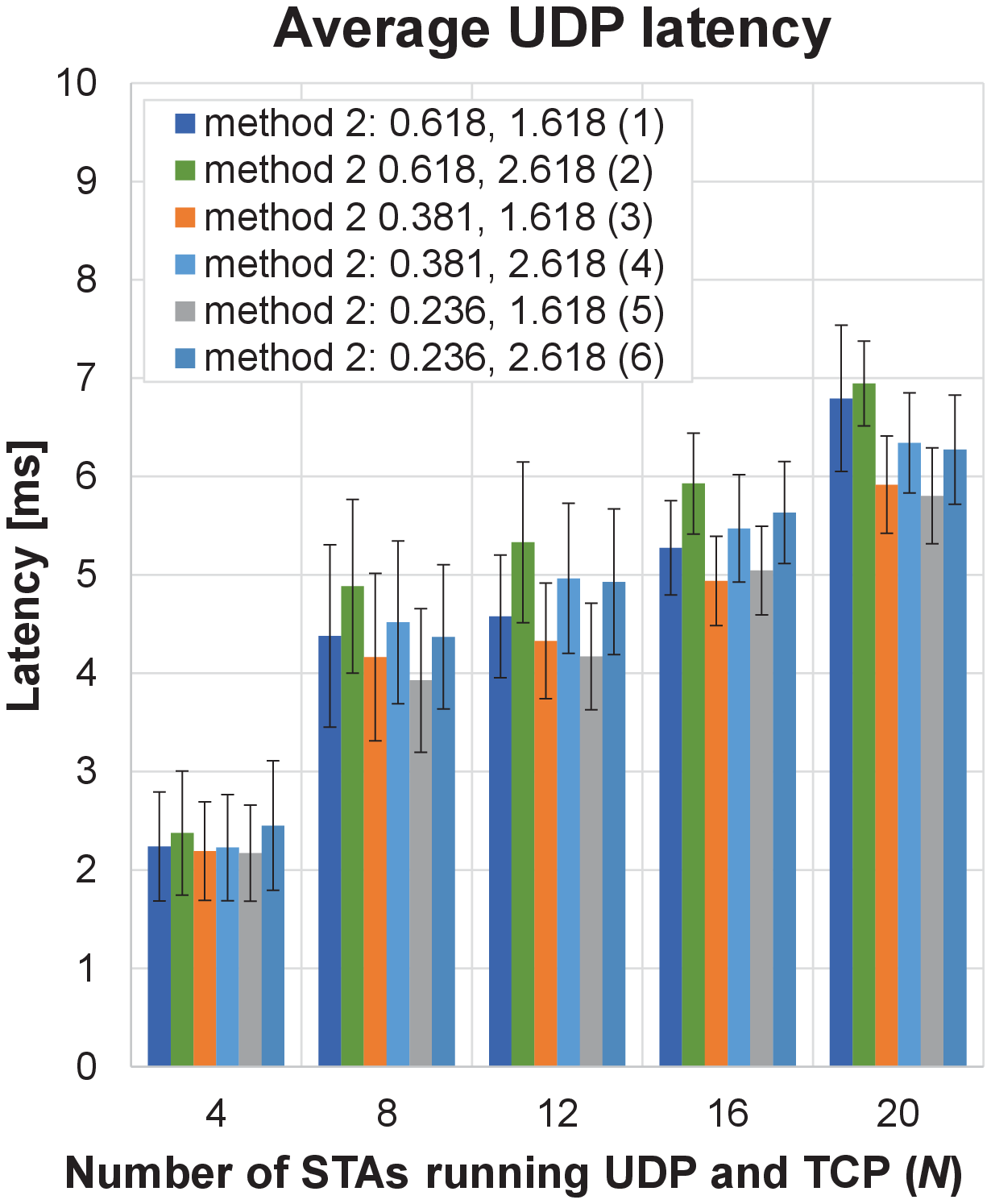}
\label{figure9b}}
\caption{Scenario with 16 APs: comparison for \textit{method} \#2 with different values of the coefficients: \textit{a)} throughput; \textit{b)} latency .}
\label{figure9}
}
\vspace*{-0.18in}
\end{figure}

Different coefficients have been employed for \textit{method} \#2, as shown in Fig. \ref{figure9}: 0.618, 0.618$^2$ (0.381) and 0.618$^3$ (0.236) for decreasing the size, and 0.618$^{-2}$ (2.618) and 0.618$^{-1}$ (1.618) for increasing it. It can be observed that an increase factor of 2.618 results in a slightly higher latency in most cases. The decreasing factor of 0.236 reduces the throughput. The pair 0.618 and 1.618 (used as default in previous tests) presents a good balance.

\section{Conclusions}
This paper has proposed the dynamic tuning of the maximum A-MPDU value in an 802.11ac WLAN. It allows to find a good balance when applications with different objectives in terms of network KPIs share the network. If aggregation is used, the throughput rises, but at the cost of a higher latency. If it is not used, or disabled, the latency gets reduced, but this comes with a penalty for the services with high throughput requisites.

The proposed algorithm allows the network manager to find an optimal balance (the ``sweet spot'') between these KPIs. In our case, a ``delay budget'' has been assigned to the real-time flows, so the algorithm has the objective of keeping the latency as close as possible to that budget, while penalizing the throughput as little as possible.

Different methods for tuning the maximum A-MPDU value have been compared. The results indicate that the use of a soft increase and decrease can reduce the latency, but it slightly penalizes the throughput. In contrast, methods with a more aggressive policy result in a higher throughput, at the cost of an increased delay.

As future work, the effect of the use of different values for the monitoring period will be studied. In addition, the dynamic tuning of the value of the ``delay budget'' depending on the end-to-end latency could be considered. If RTP traffic is used, periodic RTCP Receiver Report (RR) packets could be used for obtaining the end-to-end latency, as done in \cite{Seytnazarov}. Other policies for the tuning of the A-MPDU could also be explored, as \textit{e.g.} considering as the main objective to maximize user’s satisfaction, defined as a weighted combination of different KPIs, depending on the kind of services present in the network. As this would increase the complexity, the use of ML could be considered in that case.

\ifCLASSOPTIONcaptionsoff
  \newpage
\fi



%

\end{document}